\documentclass[twocolumn,showpacs,preprintnumbers,amsmath,amssymb]{revtex4}

\usepackage{graphicx}
\usepackage{dcolumn}
\usepackage{bm}
\usepackage{epstopdf}

\begin{document}

\preprint{}

\title{Demonstration of nonlinear-energy-spread compensation in relativistic electron bunches with corrugated structures}
\author{Feichao Fu$^{1,2}$, Rui Wang$^{1,2}$, Pengfei Zhu$^{1,2}$, Lingrong Zhao$^{1,2}$, Tao Jiang$^{1,2}$, Chao Lu$^{1,2}$, Shengguang Liu$^{1,2}$, Libin Shi$^{1,2}$, Lixin Yan$^{3}$, Haixiao Deng$^{4}$,Chao Feng$^{4}$, Qiang Gu$^{4}$, Dazhang Huang$^{4}$, Bo Liu$^{4}$, Dong Wang$^{4}$, Xingtao Wang$^{4}$, Meng Zhang$^{4}$, Zhentang Zhao$^{4}$, Gennady Stupakov$^{5}$, Dao Xiang$^{1,2}$* and Jie Zhang$^{1,2\dagger}$}
\affiliation{%
$^1$ Key Laboratory for Laser Plasmas (Ministry of Education), Department of Physics and Astronomy, Shanghai Jiao Tong University, Shanghai 200240, China \\
$^2$  IFSA Collaborative Innovation Center, Shanghai Jiao Tong University, Shanghai 200240, China \\
$^3$  Department of Engineering Physics, Tsinghua University, Beijing 100084, China \\
$^4$  Shanghai Institute of Applied Physics, Chinese Academy of Sciences, Shanghai 201800, China \\
$^5$  SLAC National Accelerator Laboratory, Menlo Park, California 94025, USA
}
\date{\today}

\begin{abstract}
High quality electron beams with flat distributions in both energy and current are critical for many accelerator-based scientific facilities such as free-electron lasers and MeV ultrafast electron diffraction and microscopes. In this Letter we report on using corrugated structures to compensate for the beam nonlinear energy chirp imprinted by the curvature of the radio-frequency field, leading to a significant reduction in beam energy spread. By using a pair of corrugated structures with orthogonal orientations, we show that the quadrupole wake fields which otherwise increase beam emittance can be effectively canceled. This work also extends the applications of corrugated structures to the low beam charge (a few pC) and low beam energy (a few MeV) regime and may have a strong impact in many accelerator-based facilities.
\end{abstract}

\maketitle

``Flat-flat'' electron beams, i.e. flat in both beam energy and current \cite{flat-flat}, are critical for achieving high performance in many accelerator-based facilities such as seeded free-electron lasers (FELs \cite{NPreview, NPsinap, SS3, FERMI, echo15}) and MeV ultrafast electron microscopes (UEMs) \cite{uem1, uem2}. By either sending the relativistic electron beams through undulators to produce coherent intense x-rays or directly using the electrons as probes, these facilities are enabling new opportunities in an extraordinarily wide array of sciences. With the beam produced in a photocathode radio-frequency (rf) gun and accelerated to high energy in rf structures, the electron beam longitudinal phase space, however, typically consists of nonlinear chirp (energy-time correlations) from the varying rf phase along the bunch. In seeded FELs, this nonlinear energy chirp deteriorates the performance of bunch compressors and broadens the FEL spectrum. In UEMs, the nonlinear energy chirp increases the beam global energy spread and thus leads to serious chromatic aberrations that reduces the microscope resolution. A similar detrimental effect also holds for ultrafast electron diffraction facilities where the nonlinear energy chirp sets the lower limit of the electron bunch length that can be achieved with velocity bunching \cite{1fs}.         

The standard method to compensate for the nonlinear energy chirp is to use a harmonic cavity (see, for example \cite{LCLSCDR}). The idea is rather simple. Consider a beam passing through two rf structures with wave numbers at $k_{1,2}$, peak energy gains at $E_{1,2}$, and phases at $\phi_{1,2}$, the energy of a particle at longitudinal position $z$ with respect to the reference particle can be written as,
\begin{equation}
E(z)=E_1 \cos(\phi_1+k_1 z)+E_2 \cos(\phi_2+k_2 z)
\end{equation}
Assuming the first rf structure is mainly used for beam acceleration and the second structure for cancellation of the nonlinear chirp, then the phase of the main acceleration structure should be correspondingly set at $\phi_1=0$ to provide maximal energy gain and the phase of the harmonic cavity should be set at the decelerating phase ($\phi_2=\pi$). Under this condition the energy chirp is canceled up to the second order if $E_2 k_2^2=E_1 k_1^2$, i.e. the required voltage for the harmonic cavity scales as $1/n^2$, where $n=k_2/k_1$ is the harmonic number.

Cancellation of the nonlinear chirp with active rf structures not only requires dedicated expensive rf stations, the relative phase of the rf structures needs to be accurately controlled as well. Another undesired outcome is that the beam energy will also be reduced by $1/n^2$ (e.g. if a C-band structure with frequency at 5712 MHz is used to cancel the nonlinear chirp from an S-band structure with frequency at 2856 MHz, the beam energy will drop by 1/4). While reduction of beam energy by a fraction for FELs is typically not a big concern, it may result in  stronger space charge effect in UEMs since the beam energy is only a few MeV \cite{uem1, uem2}. Furthermore, it may add considerable complexity and cost to such compact facilities.

Alternatively, the beam energy chirp may be canceled with passive devices such as the corrugated structures (CS) \cite{Simpson, Craievich, Antipov1, Bane12, Radiabeam, Emma, Antipov2, Deng}. This scheme exploits the interaction between the electron beam and the wake fields produced by the beam passing through the CS. By properly choosing the wavelength of the wake fields, either the linear energy chirp or nonlinear energy chirp may be ``dechirped''. In addition to greatly reduced cost and complexity compared to active compensation scheme with rf harmonic cavities, this passive compensation scheme also significantly lowers the beam energy reduction factor because the wake field wavelength can be made much shorter. Furthermore, for compact FELs driven by x-band ($\sim10$ GHz) Linac (see, e.g. \cite{Sun}) where active compensation scheme with harmonic cavity at even higher frequency becomes extremely difficult for lacking of available rf source, using CS to dechirp the beam is probably the most promising method. 

Due to these prominent benefits, almost all FEL facilities are now considering using CS to manipulate the beam longitudinal phase space to enhance the performance \cite{Deng, Gu, PAL, PSI, LCLS1, LCLS2}. In particular, most of the FEL facilities plan to use planar CS where the gap can be readily varied to change the wake field strength and wavelength to accommodate bunches with different charge and length. However, the planar geometry also excites quadrupole wake fields that have been observed to increase beam emittance by giving beam time-dependent focusing \cite{Emma}. Considering the fact that the transverse wake fields scales as $1/g^4$ ($g$ is the gap of the planar structure), this effect if not controlled, may become a limiting factor in many applications. Fortunately, theoretical analysis has shown that if the CS is composed of two identical parts with the second half rotated by 90 degrees with respect to the first half, the quadrupole wakes can be canceled \cite{quad}. 
    \begin{figure}[b]
    \includegraphics[width = 0.5\textwidth]{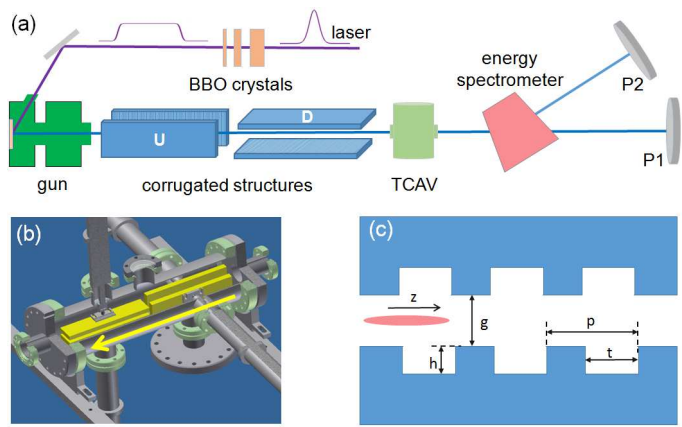}
    \caption{(a) Schematic of the nonlinear energy chirp compensation experiment (not in scale); (b) The detailed layout of the vacuum chamber where the CS is housed (the yellow arrow indicates the beam direction); (c) geometry of the planar CS (the red ellipse represents a beam propagating along the $z$ axis). 
    \label{scheme}}
    \end{figure}

Here we show that the previously observed quadrupole component of the wake fields that increase beam emittance can indeed be effectively canceled with a pair of CS with orthogonal orientations. We also show that this passive compensation scheme can be applied to near-relativistic beams (beam energy of a few MeV) with low charge (a few pC), providing important information complementary to previous theoretical and experimental efforts that focus on dechirper with ultra-relativistic beams (beam energy of a few tens of MeV and above) and high charge ($\sim$100 pC and above). 

The experiment is conducted at the Center for Ultrafast Diffraction and Microscopy at Shanghai Jiao Tong University \cite{Feichao}. The schematic layout of the experiment is shown in Fig.~1a. The electron beam is produced in an S-band photocathode rf gun, powered by a 5 MW klystron. Three $\alpha-$BBO birefringent crystals are used to shape the laser to produce a beam with nearly flat-top distribution. The beam kinetic energy at the gun exit is measured to be about 3.3 MeV with an energy spectrometer. The beam charge is about 6 pC measured with a Faraday cup. An rf transverse cavity (TCAV) is used to measure beam temporal distribution. Two 200 $\mu$m wide slits (not shown in Fig.~1a), one located upstream of the TCAV and the other just before the energy spectrometer bending magnet, can be inserted to improve the temporal and energy resolution of the measurements. Two identical aluminum planar CS with orthogonal orientations are used to compensate for the nonlinear energy chirp while effectively canceling the quadrupole wake fields. Each structure is 16 cm long and 3.2 cm wide. The location of each structure can be varied with step motors to change the gap of the structures (Fig.~1b). Specifically, the upstream CS (U in Fig.~1a) can be moved in horizontal direction and the downstream structure (D in Fig.~1a) can be moved in vertical direction. Following the notations in Fig.~1c, the corrugations are characterized by $h$=0.60 mm, $t$=0.35 mm and $p$=0.6 mm. The point charge longitudinal wake can be approximated as a damped oscillation \cite{LCLS2},
\begin{equation}
W(z)=\frac{\pi^2}{64}\frac{Z_0c}{\pi g^2}H(z)Fe^{-\frac{kz}{2Q}}\cos(kz)
\end{equation}
with $Z_0$ being the free space impedance, $c$ being the speed of light, $H(z)$ being the step function, $k=\sqrt{2p/ght}$ being the effective wave number, $F$ and $Q$ being the amplitude correction factor and quality factor which can be obtained with empirical fitting formulas.

To effectively cancel the nonlinear energy chirp, in practice a beam having a nearly flat-top distribution in the center and sharp rising and falling in the edges with FWHM duration approximately equal to half the wake field wavelength (e.g. $\lambda\approx$4.6 mm with the CS gap set at $g=3$ mm) is preferred (see, e.g. \cite{Craievich}). To achieve this, three $\alpha-$BBO crystals with temporal walkoff of 4.8 ps, 2.4 ps and 1.2 ps, are used to shape the beam temporal distribution into an approximate flat-top. This technique uses the group velocity mismatch (GVM) of the ordinary and extraordinary rays such that a laser pulse becomes two after passing though a birefringent crystal with the temporal walkoff determined by the thickness and GVM of the crystal (see, e.g. \cite{Power, Pietro, Yan}. Of particular interest here is that when the temporal walkoff is smaller than the pulse width, the laser pulses will overlap, which allows stacking the input short pulse into a long flat-top output.

The beam temporal distribution in this experiment is measured with a TCAV (an rf structure operating in TM01 mode) which gives beam a time-dependent angular kick (i.e.  $y'\propto t$) after passing through at zero-crossing phase. After a drift section the beam angular distribution is converted to spatial distribution, and the vertical axis on the phosphor screens (P1 and P2 in Fig.~1a) downstream of the TCAV becomes the time axis ($y \propto t$). The absolute time (and also the voltage of the cavity) is calibrated by scanning the rf phase and recording the vertical beam centroid motion on the screens (1 degree change in rf phase corresponds to about 1 ps change in time). The calibration coefficient is about 500 fs/mm on the phosphor screen P1. With the TCAV off, the rms vertical beam size on P1 is about 0.5 mm, corresponding to a temporal resolution of about 250 fs in this experiment.
    \begin{figure}[t]
    \includegraphics[width = 0.5\textwidth]{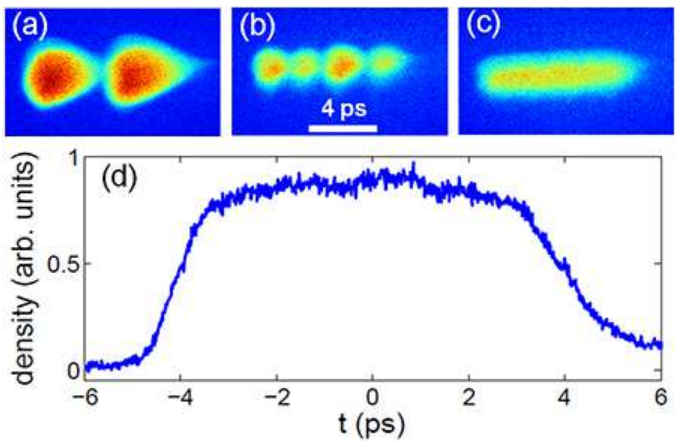}
    \caption{Time-resolved measurements of beam distribution at screen P1 with 1 crystal (a), 2 crystals (b) and 3 crystals (c) inserted (bunch head to the left); the corresponding beam current distribution with 3 crystals inserted is shown in (d).
    \label{QUAD}}
    \end{figure}

By rotating the crystals while watching the change of the beam temporal distribution, we were able to stack the Gaussian pulses in a suitable way to produce a beam with approximate flat-top distribution. The measured electron beam temporal distributions with various crystals inserted are shown in Fig~2. As can be seen in Fig.~2d, a nearly flat-top beam with FWHM of about 8 ps is produced with the 3 crystals all inserted. This allows forming a wake with suitable shape that approximates a sinusoid, critical for effective cancellation of the nonlinear energy chirp.

The beam longitudinal phase space at the gun exit is measured at the phosphor screen P2 downstream of the energy spectrometer with the TCAV on. Under this configuration the vertical axis on P2 becomes the time axis and the horizontal axis becomes the energy coordinate. The measured beam longitudinal phase space with the two structures open (e.g. $g$=30 mm) is shown in Fig.~3a. Due to space charge effect, the nonlinearity of the energy chirp slightly deviates from the curvature of the rf field. Specifically, because the longitudinal space charge force pushes the beam in the head further forward, the energy of the particles in the bunch head is increased compared to the space charge free case and thus a ``kink'' is formed in the bunch head. Also the chirp in the bunch head is larger than that in the bunch tail, as also seen in simulations (not shown). When the gap of the upstream CS is reduced to $g$=3 mm, the longitudinal wake field partially cancels the nonlinear energy chirp (Fig.~3b). Finally when the gap of the downstream CS is also reduced to $g$=3 mm, the longitudinal phase space stands upright with the nonlinear chirp greatly canceled (Fig.~3c). 
    \begin{figure}[b]
    \includegraphics[width = 0.5\textwidth]{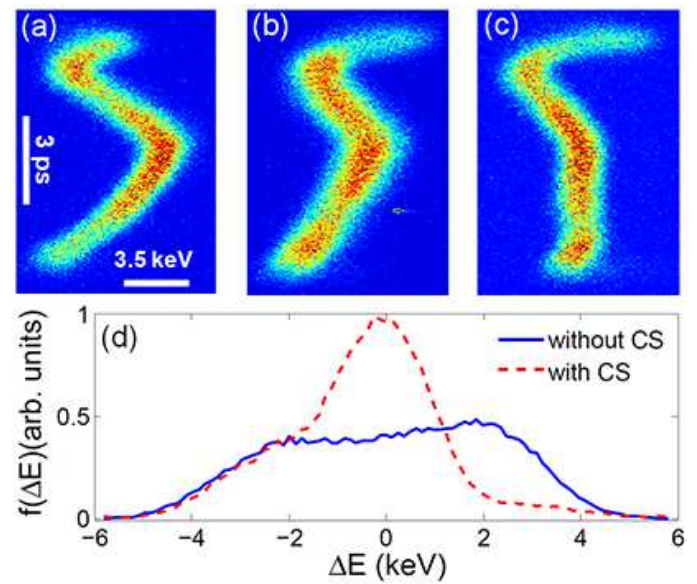}
    \caption{Time-resolved measurements of beam longitudinal phase space distributions (bunch head to the up) at screen P2 with the two structures open (a), with one of the CS gap reduced to 3 mm (b) and with the gaps of the two CS both set at 3 mm (c); the corresponding projected beam energy distribution (d).  
    \label{QUAD}}
    \end{figure}

The projected energy distributions for Fig.~3a and Fig.~3c are shown in Fig.~3d, where one can see that the FWHM beam energy spread has been reduced from about 6.7 keV (dominated by nonlinear energy chirp) to about 2.4 keV (dominated by slice energy spread and resolution of the energy spectrometer) with the two CS. It should be pointed out that in our experiment the longitudinal space charge force modifies the curvature of the phase space such that the wake field can only compensate for the nonlinear energy chirp for part of the beam. In particular, the chirp in the very front of the bunch which contributes to a few percent of the beam charge, is actually further increased by the CS wake because it has the wrong sign (Fig.~3c). Nevertheless, with the nonlinear energy chirp for the main part of the beam effectively compensated, the beam global energy spread is only limited by the slice energy spread. 

It should be noted that the beam slice energy spread increases after passing through a TCAV \cite{TCAV1,TCAV2}, which makes the measured beam energy spread at screen P2 larger than that at the exit of the CS (in simulation the beam slice energy spread before the TCAV is about 0.2 keV). The beam energy spread growth is from the longitudinal electric field which varies linearly with transverse distance. With the 200 $\mu$m slit upstream of the TCAV inserted and the voltage of the TCAV measured to be about 160 kV, the TCAV induced beam energy spread is estimated to be about 1.9 keV (FWHM), similar to that measured at P2. Furthermore, the resolution of the energy spectrometer is about 1.5 keV, not sufficient to resolve the slice energy spread at sub-keV level. Therefore, the nearly 3 fold reduction (Fig.~3d) should be considered as the lower limit and the true energy spread suppression factor at the exit of the CS is likely to be much larger.  
    \begin{figure}[t]
    \includegraphics[width = 0.49\textwidth]{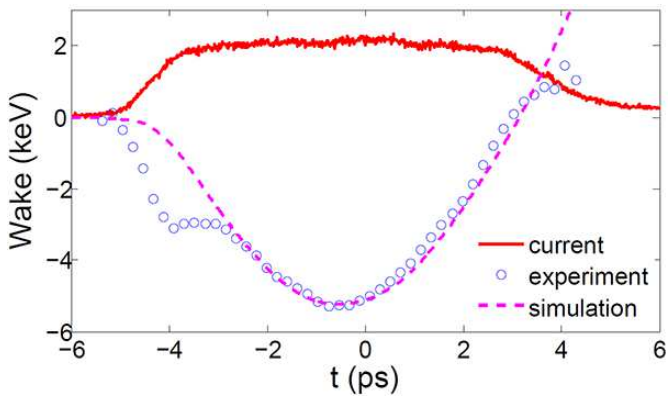}
    \caption{(a) Measured and simulated wake potential of the CS. The bunch distribution, with the head to the left, is also shown with red line.  
    \label{scheme}}
    \end{figure}

The longitudinal wake produced by the CS is shown in Fig.~4 (blue circles), which is obtained by calculating the energy differences between Fig.~3a and Fig.~3c as a function of longitudinal position. The simulated wake (magenta dashed line in Fig.~4) calculated by convolving the beam longitudinal distribution with the point charge wake (Eq.~(2)), is in reasonably good agreement with measurement. Note, in Fig.~4 only the wake for the main part of the beam is shown and in the simulation the beam charge is taken to be 7 pC slightly larger than the measured beam charge (6 pC). The wake for the very front and tail of the beam is not accurately quantified because of the low signal to noise ratio from the limited number of particles in these regions. It is worth mentioning that compared to using C-band harmonic cavity ($n=2$) to remove the nonlinear energy chirp where the beam energy would drop by about 800 keV (25\% of the beam kinetic energy), here by using corrugated structure to provide wake fields at much higher frequency ($n\approx23$) the beam energy only reduces by about 6 keV (0.2\% of the beam kinetic energy, as can be seen in Fig.~4). Also this scheme does not require dedicated equipment to control the phase of the wake field, as needed in active compensation scheme.  
    \begin{figure}[t]
    \includegraphics[width = 0.5\textwidth]{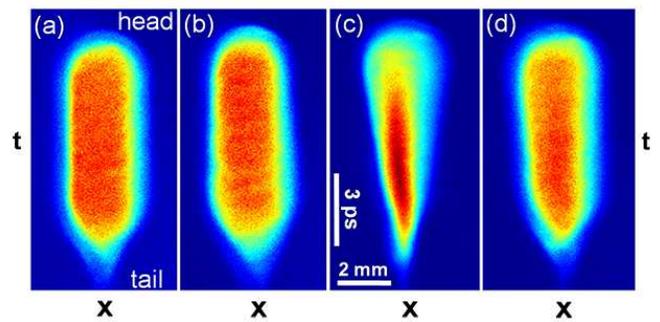}
    \caption{Time-resolved measurements of beam distribution at screen P1, showing no quadrupole wake when the gaps of the two structures are open (a), time-dependent defocusing in $x$ from the quadrupole wake of the upstream CS (b), time-dependent focusing in $x$ from the quadrupole wake of the downstream CS (c), and again almost no quadrupole effect with the gaps of the two structures both set at 3 mm (d). 
    \label{QUAD}}
    \end{figure}

In addition to removing the beam nonlinear energy chirp, we also show how the associated quadrupole wake fields of a pair of planar CS with orthogonal orientations cancel each other. This measurement is done at the screen P1 with the TCAV on. The streaked beam distribution with the two CS widely open is shown in Fig.~5a. With the gap of the upstream CS reduced to 3 mm, the beam distribution is shown in Fig.~5b, where one can see that the wake field produced by the head of the beam defocuses the tail of the beam such that the bunch tail horizontal size is slightly increased. This quadrupole component is best seen with the gap of the downstream structure reduced to 3 mm (the upstream structure is open). As shown in Fig.~5c, the time-dependent quadrupole wake field strongly focuses the tail part of the beam. It should be noted that because the beam has relatively large horizontal size at P1, the beam size is more sensitive to focusing than defocusing. With the gap of the two CS both set at 3 mm, the quadrupole wake fields cancel each other just as that predicted in theory\cite{quad}, resulting in a relatively uniform beam (Fig.~5d), similar to that without CS.

In summary, we have presented the measurement of beam phase space manipulation with CS in the low beam energy (a few MeV) and low beam charge regime (a few pC), providing important complementary information to previous world-wide efforts that focus on beams with high energy ($\sim$100 MeV and above) and high charge ($\sim$100 pC and above). In addition to directly showing the compensation of the beam nonlinear energy chirp with CS through measurements of the beam longitudinal phase space, we also demonstrate for the first time that the quadrupole component of the wake fields that increases beam emittance can be effectively canceled with a pair of planar CS with orthogonal orientations. The results are in good agreement with simulations and should forward the applications of this technique in simplifying the design and enhancing the performance of many accelerator-based scientific facilities.  

We thank J. Chen and Y. Chen for maintenance of the laser system and Y. Du for useful discussions. This work was supported by the Major State Basic Research Development Program of China (Grants No. 2015CB859700 and 2011CB808300)) and by the National Natural Science Foundation of China (Grants No. 11327902, 11475097, 11275253) and by the U.S. DOE under Contract No. DE-AC02-76SF00515. One of the authors (DX) would like to thank the support from the Program for Professor of Special Appointment (Eastern Scholar) at Shanghai Institutions of Higher Learning (No.SHDP201507).

*~dxiang@sjtu.edu.cn

$\dagger$~jzhang1@sjtu.edu.cn

\end{document}